\documentclass{article}

\def\np#1#2#3{Nucl. Phys. {\bf{B#1}} (#2) #3}
\def\pl#1#2#3{Phys. Lett. {\bf{B#1}} (#2) #3}
\def\ij#1#2#3{Int. J. Mod. Phys. {\bf{A#1}} (#2) #3}
\def\mp#1#2#3{Mod. Phys. Lett. {\bf{A#1}} (#2) #3}
\def\M{M}

\hoffset= - 1.0in         

\begin{document}

\hfill ITEP-TH-04/05

\bigskip

\centerline{\Large{
Challenges of Matrix Models
}}

\bigskip

\centerline{A.Morozov}

\centerline{\it ITEP, Moscow, Russia}

\bigskip

\centerline{ABSTRACT}

\bigskip

\centerline{Brief review of concepts and unsolved
problems in the theory of matrix models.}
\centerline{Contribution to Proceedings of Cargese 2004.}

\bigskip

\bigskip

\section{Introduction}

Matrix models appear again and again at the front line of theoretical
physics, and every new generation of scientists discovers something
new in this seemingly simple subject. It is getting more and more
obvious that matrix models capture the very essence of general
quantum field theory and provide the crucial representative example
for the {\bf string theory} \cite{ufn2}:
matrix models play for string theory the role, which
harmonic oscillator plays for quantum mechanics.
Essential difference is, however, that we have a nearly exhaustive
understanding of harmonic oscillator, while nothing like
complete description of matrix models is yet available.
Probably, the time is coming to begin a systematic analysis
with the goal of building up a theory of matrix model partition
functions as the first special functions of string theory \cite{amm1}.
Various applications of matrix model techniques should use
these functions as building blocks for formulation of their
results, thus separating the physical content of different applications
from common mathematical formalism. The goal can be to build up an
analogue of the powerful free-field formalism, developed in 1980's,
which allowed to reduce many problems in {\bf perturbative
string theory} \cite{MP}
(and related issues in representation theory of
Kac-Moody algebras, $2d$ conformal field theory and finite-zone solutions
in integrable-systems theory) to almost classical set of special
functions: Riemann's theta-functions \cite{Mum,Fay},
associated with Riemann surfaces.
The problem is that matrix model partition functions
are non-trivial generalization of Riemann's theta-functions,
which were not fully investigated in XIX-th century, therefore
the theory should involve essential new ideas, both in physics and
in mathematics. This is the reason why the progress in this
field -- actually known under the nick-name of {\bf non-perturbative
string theory} -- is considerably slower. More and more people
begin to realize that generic problems of non-perturbative string
theory, i.e. the general theory of phase transitions,
have a lot in common, and these most interesting common properties
are clearly represented  at the simplest level of matrix models.
Complications, introduced by sophisticated context of particular
applications can and should be separated from important content,
captured at the matrix model level.

In development of matrix model theory
one can distinguish between several stages. In fact, these stages
are typical for every chapter of the string theory,
and they characterize not so much the history, but the
kind of questions that are posed and the level
of abstraction in analysis of these questions.
Whenever a new problem arises (or is re-addressed again),
it is quite useful to understand its place in this hierarchy.

\section{Classical period: introduction of models}

During the {\it classical period} the main task was to
{\it study} concrete phenomena, involving matrix models.
Since in these notes we are not going to discuss applications,
we just mention a few crucial theoretical methods, developed
at this stage.

The most important idea was to reduce the $N^2$-fold
matrix integrals like
\begin{equation}
\label{mamo}
Z(t|N) = \frac{1}{{\rm Vol}(U(N))}
\int_{N\times N} d\Phi \exp \left(\sum_k t_k{\rm Tr}\ \Phi^k\right)
\end{equation}
to $N$-fold integrals over eigenvalues $\phi_i$
of matrix-valued fields $\Phi = U^\dagger {\rm diag} (\phi_i) U$.
This idea, closely related in general context to the
problem of gauge invariance, is technically based on
the possibility to explicitly integrate over "angular variables"
in the simplest situations, of which (\ref{mamo}) is a
representative example. The single known exactly solvable
generalization, which goes slightly beyond this example,
involves the Itzykson-Zuber integral \cite{IZ}
\begin{equation}
\label{IZ}
\int_{N\times N} [dU] \exp ({\rm Tr} AU^\dagger BU),
\end{equation}
which is used in construction of two important classes of
matrix theories: generalized Kontsevich model \cite{Ko,GKM},
\begin{equation}
\label{GKM}
Z_{GKM}(L|W) \sim
\int_{m\times m} dM \exp \left[{\rm tr} \left(W(M) + LM\right)\right]
\end{equation}
and even further generalized Kazakov-Migdal-Kontsevich model
\cite{KM,GKMMM},
\begin{eqnarray}
\label{GKMKM}
Z_{G(KM)^2}(\{L_\mu\}|\{W_\alpha\}) \sim
\prod_{\mu,\alpha}\int_{m\times m}
dM^{(\alpha)} [dU^{(\alpha\beta)}][dU^{(\alpha\mu)}]\cdot
\nonumber \\
\cdot\exp \left[{\rm tr} \left( W_\alpha(M_{\alpha}) +
M_\alpha U_{\alpha\beta}^\dagger M_\beta U_{\alpha\beta} +
L_\mu U_{\alpha\mu}^\dagger M_\alpha U_{\alpha\mu} \right)
\right],
\end{eqnarray}
defined for any graph with Hermitian fields
$M_\alpha$ and background fields $L_\mu$
standing in the vertices and unitary fields $U_{\alpha\beta}$
and $U_{\alpha\mu}$ on the links.
In more general situations (sometime quite important for applications,
e.g. involving non-trivial multi-brane backgrounds), angular
integration is highly non-trivial and remains an open problem.
The simplest of such difficulties are echoed in the Gribov
copies problem in Yang-Mills theory, while treatment of
more sophisticated unitary-matrix integrals face problems,
typical for adequate treatment of quantum gravity.
The class of theories where gauge (angular) variables
can be integrated out in effective way is referred to as
{\bf eigenvalue models} \cite{ufn3}.
Most of further development
concerns eigenvalue models, which until recently remained
implicitly synonymous to matrix models.

The next idea of the {\it classical period} used the fact,
that angular integration usually provides non-trivial, but
very special measures on the space of eigenvalues,
which is often made from Van-der-Monde determinants \cite{VdM}.
For example, (\ref{mamo}) can be transformed into
\begin{equation}
\label{eigenvrep}
Z(t|N) = \prod_{i=1}^N \int d\phi_i
\exp \left(\sum_k t_k\phi_i^k \right)
\prod_{i<j}^N (\phi_i-\phi_j)^2,
\end{equation}
where $\prod_{i<j}^N (\phi_i-\phi_j)^2 =
\left(\det_{i,j} \phi_i^{j-1}\right)^2$
can be also considered as discriminant of auxiliary polynomial
$P_N(z) = \prod_{i=1}^N (z-\phi_i)$.
Occurrence of determinants suggested two alternative technical
methods of investigation of eigenvalue models:
technique of orthogonal polynomials \cite{orthopol} and
free-fermion representation \cite{orthopol,freefer}.
Occurrence of discriminants
was not fully exploited yet, though it opens a set of interesting
possibilities, both technically and conceptually.

The last idea of the {\it classical period} which needs
to be mentioned, is the idea of {\bf continuum limit}, when
the matrix size $N$ tends to infinity. In fact there are
infinitely many different continuum limits -- relevant for
different particular applications, -- and only very few have
been analyzed so far. However, from the very beginning the
main fact was broadly realized: continuum limits of matrix
models posses description in terms of Riemann surfaces
\cite{loope,contlimRS}. Today we know that occurence of spectral
surfaces is a general phenomenon, and they occur already at finite
values of $N$ \cite{amm1}.

\section{First stringy period: generalizations and
hunt for structures}

The {\it first stringy period} is characterized by the
change of interests: from tools to theory.
Instead of studying various applications
and developing adequate technical methods to answer the
problems, posed by these applications, attention gets concentrated
on the search and understanding of internal structures.
Instead of {\it "study"} the main slogan becomes
{\it "deform and generalize"} -- this is the standard string
theory method of revealing hidden structures.
It is at this stage that the three main
inter-related structures were discovered
behind eigenvalue matrix models: rich Ward identities,
integrability and CFT representations.

\subsection{Ward identities}

Occurence of {\bf Ward identities} is the pertinent property of
every integral (and thus of {\it quantum mechanics} and all its
generalizations, like quantum field and string theory):
they reflect invariance of the integral under the change of
integration variable -- an archetypical example of auxiliary
field. However, this obvious hidden
symmetry manifests itself in a rather sophisticated manner:
as relation between various correlators in the theory.
String theory normally deals with partition functions: the
generating functions of all the correlators, summed up with the
coefficients like $t_k$ in (\ref{mamo}),
which have the meaning of extra coupling constants,
and can be considered as providing the deformation of the
bare action. This formalism allows
to treat Ward identities as {\it equations} for partition function,
because the change of integration variables can be compensated
by the change of the coupling constants, if there are many enough
\cite{ufn3}. In particular case of the integral (\ref{mamo})
the equations are known as {\bf Virasoro constraints} \cite{virco}
or {\bf loop equations} \cite{loope},
\begin{equation}
\label{virco}
\hat L_-(z) Z(t|N) = 0,
\end{equation}
$$\hat L_-(z) = \sum_{m\geq -1}z^{-m-2}\hat L_m =
{\cal P}_-\left(
\sum_{m=-\infty}^\infty z^{-m-2}\hat L_m\right) =
{\cal P}_-\left(\frac{(\partial\hat\phi(z))^2}{2}\right)$$
$$\hat\phi(z) = \sum_{k\geq 0}\left(
t_kz^{k} - \frac{1}{2kz^{k}}\frac{\partial}{\partial t_k}
\right)$$
For the integral (\ref{GKM}) the Ward identities can be
written in two forms: either as a Gross-Newman equation
\cite{GN},
\begin{equation}
\label{GN}
\left(W'\left(\partial/\partial L_{tr}\right) + L\right)
Z_{GKM}(L|W) = 0,
\end{equation}
which generalizes a similar equation \cite{ufn3}
for the Itzykson-Zuber integral (\ref{IZ}),
or as a set of peculiar ${\cal W}_n$-constraints (where $n =
{\rm deg}\ W'(z)$) \cite{Wcon}. A particular case of these
equations in the case of $n=2$ is
{\bf continuum Virasoro constraint} \cite{DVV,MMM},
\begin{eqnarray}
\label{contvirco}
\hat{\cal L}_{2m} Z_{GKM}\left(L\left|\frac{1}{3}M^3\right)\right.
= 0, \ \ m\geq -1, \nonumber \\
\hat{\cal L}_{2m} = \frac{1}{2}\sum_{{\rm odd}\ k=1}^\infty
kt_k\frac{\partial}{\partial t_{k+2m}} +
\frac{1}{4}\sum_{{\rm odd}\ k=1}^{2m-1}
\frac{\partial^2}{\partial t_k\partial t_{2m-k}} +
\frac{1}{16}\delta_{m,0} + \frac{1}{4}t_1^2\delta_{m,-1}
\end{eqnarray}
where $t_k = r_k + \frac{1}{k}{\rm tr} L^{-k/n}$ and
$r_k = \frac{n}{k(n-k)}{\rm res}\left(W'(z)\right)^{1-k/n}dz =
-\frac{2}{3}\delta_{k,3}$.
See \cite{ufn3} for detailed discussion of Ward identities for
eigenvalue models and \cite{KMMMMPLA} for more technicalities,
related to generalized Kontsevich model.

\subsection{Integrability and RG evolution}

{\bf Integrability} of matrix models
means that partition functions satisfy not only linear equations
(Ward identities), but also bilinear Hirota-type equations,
$$
\Delta (Z\otimes Z) = 0.
$$
Technically, the proofs rely upon determinant representations
of eigenvalue models, which are, in turn, immediate corollaries
of determinant structures of integration measures, see
\cite{ufn3} and references therein. However, the true meaning
of integrability remains obscure. Several inter-related ideas should
be somehow unified to clarify the issue.

First, integrability
should express the fact that the system of Ward identities is rich
enough: enough to specify the partition function almost
unambiguously -- up to some easily controllable degrees of
freedom, a sort of {\it zero modes} with some clear cohomological
interpretation. See \cite{mirmor} for the relevant notions of
{\it strong} and {\it week completeness}.

Second, bilinear equations normally involve
the matrix size $N$ in non-trivial way: for example, the simplest
bilinear equation for partition function (\ref{mamo}) --
the lowest term of the Toda chain hierarchy -- states \cite{GMMMO}
that
\begin{equation}
\label{firstToda}
Z_N\frac{\partial^2 Z_N}{\partial t_1^2} -
\left(\frac{\partial Z_N}{\partial t_1}\right)^2 =
Z_{N+1}Z_{N-1}
\end{equation}
This means that bilinear equations involve not only
variations of the coupling constant $t_k$, but also those of
conjugate variables, like $N$. However, in (\ref{mamo})
only $N$ -- a conjugate variable for $t_0$ (in the sense that
$\partial \log Z_N/\partial t_0 = N$) -- is present,
while there is nothing like conjugate parameters for all
other $t$'s. This can
suggest that Hirota equations for such restricted partition
function are too special to reveal their general structure.
In particular, (\ref{firstToda}) holds only for the simplest
phase of the model (\ref{mamo}), other phases belong to
multi-component generalizations of Toda chain hierarchy.

Third, while coupling constants parameterize the {\bf bare action},
i.e. the weight of summation over paths in functional integral,
the conjugate variables should rather characterize the integration
domain (range of integration, boundary conditions, the target space
-- whatever formalism one prefers to use in introducing them). In this
sense conjugate variables are intimately related to generalized
renormalization group flows, which are supposed to do exactly the same:
describe the dependence
of functional integral on the integration domain.

Relation between {\bf integrability} and {\bf renormalization group}
is one of the main open problems in modern theoretical physics
\cite{ufn2,ufn3,Intre}.
To get fully related, both these concepts should be considerably
modified. To become a pertinent feature of all partition functions,
integrability should not be restricted to ordinary $\tau$-functions
\cite{JM},
associated with free fermions and single-loop Kac-Moody algebras
at level $k=1$. Indeed, $\tau$-functions and bilinear Hirota-like
equations can be defined for arbitrary Kac-Moody and even more
general Lie algebras and quantum groups \cite{KhL}.
However, the theory of such $\tau$-functions is not reducible to
that of  Plucker (free-fermion) determinants and requires the
full-scale application of the free-field formalism \cite{MP}
and more sophisticated determinant formulas. Expressions of this type
are expected to arise in the study of unitary matrix integrals,
but only first steps are being done in this direction
\cite{uniin1,uniinl}.

Renormalization group theory also requires considerable generalizations,
of which we put especially emphasize on two.
First, as already mentioned, it should allow arbitrary variations
of integration domain, not just a one-parameter cut-off procedure:
renormalization group should study the changes of the {\it shape}
of integration domain, not just of its {\it volume}. This seems to
be already a widely accepted generalization.
Second, renormalization group theory should be made applicable
to the study of fractal structures, which often reproduce themselves
(exactly or approximately) at different scales. This means that
the theory should allow non-trivial, periodic and perhaps even
chaotic, renormalization group flows \cite{MN,charg}, and this should
not contradict the obvious uni-directional nature of these flows,
generated by integrating out degrees of freedom.
This entropy-like feature of renormalization group is usually
expressed through the {\bf $c$-theorem} \cite{cth}, but
in the case of systems with infinitely many degrees of freedom
the $c$-function
can easily have angular nature, and monotonic decrease of $c$ does not
contradict the existence of periodic motions. Besides the obvious
example of self-similar fractals, today we already have some
field-theory examples of such behaviour \cite{perrg} -- but only
at the level of {\it flows}.
It is not so easy to provide examples of {\it partition functions}
(not just flows) with
such properties, and the reason for this is simple: we usually consider
as {\it solvable} the systems where answers can be expressed through
conventional special functions, and these -- almost by definition --
have oversimplified branching structure. We do not posses any language
to describe more general -- and more realistic situations. At the
same time such language should exist, because what seems to be
sophisticated (fractal and/or chaotic) phase structure is in fact a
highly ordered and pure algebraic pattern \cite{DM}, not so much
different from the algebraic-geometry background under the theory of
conventional special functions.
The jump between {\it order} (integrability) and {\it chaos}
(infinite-tree structure of bifurcations or phase transitions)
can be not so big, and most probably the careful study of partition
functions already at the matrix-model level will reveal the deep
inter-relation (duality) between the two concepts.

This duality should play a big role in {\bf landscape theory},
which studies the distributions of various algebro-geometric
quantities on moduli spaces (either of coupling constants or on the dual
space of different branches) and their interplay with renormalization
group flows. The crucial hidden double-loop structure, which is supposed
to become the basis of such theory \cite{GLM}, should be seen already
at the level of matrix models. For old and new attempts to apply --
the yet undeveloped -- landscape theory to phenomenology see \cite{ufn2}
and \cite{lands} respectively.

\subsection{Relation to conformal theory}

The third structure behind eigenvalue matrix models is that of
$2d$ conformal field theories: partition functions of matrix models
coincide with certain correlators in conformal models. This fact
is of course related with the free-fermion and KP $\tau$-function
representations of simplest partition functions, but it was quickly
realized to have a more general meaning \cite{confmamo}:
CFT representations are also closely related to Ward identities.
In the simplest example of (\ref{virco}) the natural observation is
that $\frac{1}{2}(\partial\hat\phi)^2(z)$ looks like the stress
tensor $T(z)$ of the $2d$ free scalar field, and therefore one can
look for an intertwiner, relating operator $\hat L(z)$ in (\ref{virco})
with $T(z) = \frac{1}{2}(\partial\varphi)^2(z)$ in the free field
theory. The intertwiner in this case is just
$$
\langle 0 | e^{\oint v(x)\partial\varphi(x)},
$$
where $\varphi(z)$ is the $2d$ scalar, $\langle 0|$ -- the left
vacuum, annihilated by $\varphi_+(z)$, $\langle 0 |\varphi_+(z) = 0$,
and $v(z) = \sum_k t_kz^k$. Obviously,
$$
\hat L(z) \langle 0 | e^{\oint v(x)\partial\varphi(x)} =
\langle 0 | e^{\oint v(x)\partial\varphi(x)}
\frac{1}{2}(\partial\varphi)^2(z),
$$
and acting on the right vacuum $(\partial\varphi)^2_-(z)|0\rangle = 0$,
so that
\begin{equation}
\label{cftrep}
\hat L_-(z)\langle 0 | e^{\oint v(x)\partial\varphi(x)} Q |0\rangle = 0.
\end{equation}
Here $Q$ is inserted in order to make matrix element non-trivial,
and the only requirement is that $Q$ commutes with $T(z) =
\frac{1}{2}(\partial\varphi)^2(z)$. Such $Q$ can be made out of the
{\it screening charges} $Q_\pm = \oint e^{\pm\sqrt{2}\varphi(x)}$,
and $Z(t|N)$ in (\ref{mamo}) is reproduced if $Q = Q_+^N$.
Many generalizations are straightforward, among them the important
class of quiver matrix models \cite{confmamo,Kostov1}.
An immediate desire is to find a representation of this kind for
Kontsevich $\tau$-function -- solution of continuum Virasoro
constraint (\ref{contvirco}), but this problem appears surprisingly
complicated and still remains open, see \cite{Moore,Kostov2}
for important but still insufficient steps towards its solution.

\section{Transcendental period: absolutization of structures}

The {\it transcendental} or {\it second stringy} period is characterized
by {\it absolutization of structures}, revealed at the previous stage.
This means that the logic is inverted: now the {\it structures} are given,
and we look for generic objects, possessing such structures --
expecting in advance that some of these objects can appear different
from original matrix integrals. Today we are mostly at this stage,
at least in the theory of matrix models, and only the first careful
steps are being done in attempts to understand the hidden meaning
of the structures that we observed in applications --
and continue to observe again and again in new physical systems when they
come to our attention.

Discussion of generic CFT correlators as well as of generic
$\tau$-functions is -- old and important -- but still not a very
constructive problem. It lies at the very core of landscape theory,
but existing theoretical methods are too week to address the problem
directly. The structure which can be rather effectively analyzed by
available tools or by their straightforward modifications is that
of Ward identities. Working on this structure one can also hope to
develop stronger methods, applicable to analysis of the other two
structures. To mention just one open question, illustrating the degree
of ignorance in this field, it remains unclear
to what extent the Virasoro constraints (\ref{virco}) and
(\ref{contvirco}) {\it per se}, without explicit integral
representations like (\ref{mamo}), imply integrable structure of
partition function.

\subsection{Ward identities}

Defining partition function as solution to Ward identities is
a typical $D$-module-style definition, i.e. basically
a problem from {\it linear algebra}, since Ward identities
are {\it linear} equations (this is what makes examination of
this structure much simpler than integrable one, related to
quadratic equations). Since we do not expect any mysteries in
linear algebra (or, better to say, we think that we already
know all of them), the theory building is straightforward, but
by no means trivial: we know what to search for, but exhaustive
and explicit solution can be quite tedious. We are not yet
at the stage when solution of a linear-algebra problem,
especially infinite-dimensional, and all its properties can be
predicted {\it a priori}.

Normally, the set of questions
one asks about the solution consists of several groups.

\subsubsection{Domain of definition}

First of all, one should decide what class of functions one
wants the solution to belong to. Basically, there are two
principal alternatives: formal series and globally defined
functions. However, even if one chooses formal series, it is
still necessary to specify, where {\it singularities}
are allowed to occur.
The simplest alternative is between series in positive
powers of a variable (singularities are allowed at infinity
of the Riemann sphere) and in all integer powers (singularities
allowed at infinity and at zero). However, there are many
more possibilities: singularities can be allowed at other
points (then negative powers of $[z-a]$ can be allowed),
the underlying {\it bare} Riemann surface need not be
a sphere (then certain fractional powers of $[z-a]$ are allowed)
etc. More than this, different
variables can have different kinds of singularities and we
have infinitely many such variables. In different
applications different requirements are imposed, and one and
the same quantity -- unique partition function = generic solution
to Ward identities -- can show up in absolutely different way:
different {\bf branches} are relevant for different applications.
What can look like a singularity from the point of view of
particular application, can be nothing but a branching point
or {\bf phase transition} to another branch, which acquires a
natural interpretation in terms of an absolutely different
application. Once again this emphasizes that the problem of
partition functions -- even one particular partition function,
one particular matrix model $D$-module,-- is almost
undistinguishable from the central problem of the string theory
\cite{ufn2}, which can be formulated as a search of unique
universal partition function (a {\it universal object}
of quantum field theory), of which all other thinkable partition
functions are particular branches and sub-families.
Returning closer to the Earth, even speaking about formal series
one -- explicitly or implicitly -- imposes certain requirements
on allowed singularities of solutions, which often do not follow
from the Ward identities themselves. Actually, the
"rich" Ward identities, that we spoke about in the previous
section, fix everything but the behaviour at singularities,
they are rich enough to fully constrain the dependence
on the choice of the action (on coupling constants),
but not rich enough to fix conjugate dependencies, e.g.
boundary conditions at possible singularities:
in certain sense they fix exactly one
half of possible freedom. This is a typical thing for the
equation on a wave function in quantum mechanics to do
(constrain $q$-dependence but ignore $p$, or in other words,
there are many solutions of the same Shroedinger
equation, differing, say, by the values of energy):
this observation is a starting point on the
road, leading to identification of Ward identities with
Shroedinger-type equation and the partition function with
a (string-field theory) wave function.

Convenient way to speak about emerging freedom is in
terms of "globally defined" functions: to fix it one can ask
to solve Ward identities
for functions on Riemann sphere or on torus or anywhere
else. The problem is that in most interesting cases -- and
matrix model partition functions are not an exception --
singularities are unavoidable. Moreover, small relaxations
of compactness, like "functions with simple poles", which
were enough for the formalism of free fields (and thus in
dealing with arbitrary $2d$ conformal models), now are not
sufficient: $\tau$-functions are different from conformal
blocks because they have essential singularities, even if
we take those defined on punctured Riemann surfaces of
finite genus. Actually, things are even worse: interesting
partition functions require this genus to be infinite, if
at all defined. Therefore this language -- though widely used --
by itself does not provide unambiguous classification of solutions.

\subsubsection{Generating functions}

The next question -- after the set of allowed functions is
somehow specified -- is  how to represent the answer.
There are few chances that the full answer will be any known
special function: as already stated, the most important
thing is that matrix models produce {\it new} special
functions.  What is important, however, these new functions
are not {\it too} new, they are just one step forward
more involved than the known
special functions, namely the Riemann's theta-functions,
which occupy the previous level of complexity. We expect
that all kinds of simplifications of matrix model partition
functions reduce them to quantities, already expressible through
Riemann's theta functions. What can be these simplifications?
Instead of considering a generating function of {\it all}
correlators, one can select some one-parametric family,
and such reduced generating function can be a candidate.
One can take one or another limit (say, one or another
continuum limit: naive, double-scaling, fixed genus, ...).
One can take a ratio of different branches (a kind of
monodromy matrix) or some more sophisticated combination
of those.

\subsubsection{Different realizations}

As usual for a $D$-module, one can look for solutions
of linear equations in different forms.

Integral formulas like (\ref{mamo}) are particular
examples of one possible -- integral -- representation
(or {\it realization}, to avoid confusion with group-theory
representations) of solutions. For integral realizations
the Ward identities play a role of Picard-Fucks equations:
solutions are periods of the form which is converted into
a full derivative by application of the corresponding
operator (which generates Ward identity). Of course, in such
situation one can integrate along any closed contour and
different non-homological contours give rise to different
solutions. Therefore, when one calls eq.(\ref{mamo}) Hermitian
one matrix model, "Hermitian" actually refers to the
measure, $d\Phi = \prod_{i,j=1}^N d\Phi_{ij}$ (dictated,
in its turn, by the norm $||\delta \Phi||^2 = {\rm Tr}
\ \delta\Phi^2$), but not to the integration contour: when
one defines partition function as solution of the Virasoro
constraints, there is no reason to integrate in (\ref{mamo})
over the contour $\Phi_{ij}^* = \Phi_{ji}$,
associated with Hermitian matrices. Moreover, since
$D$-module is defined by linear equations, the superposition
principle is applicable, and a sum of any two solutions is
again a solution. If matrix integrals like (\ref{mamo}) with
different $N$ satisfy the same Ward identities like
Virasoro constraints (\ref{virco}), then linear combinations
of integrals with different $N$ still are solutions. This
means that the matrix size $N$ can be also interpreted as
characteristic of integration contour, moreover, this
characteristic is not obligatory a positive integer, but
can be also negative, rational and even complex-valued.

Differences between possible integral realizations of
one and the same partition function are not exhausted
by freedom in the choice of integration contours, even
with extended interpretation of the term "contour".
Partition function can be represented by absolutely different
classes of matrix models. The simplest important example is
provided by the model (\ref{mamo}):
it can be represented not only by (\ref{mamo}), but also
by a Gaussian integral \cite{ChMak} from Kontsevich family
(\ref{GKM}) with $t_k = \frac{1}{k}{\rm tr}L^{-k}$
\cite{KMMMToda,ufn3}:
\begin{eqnarray}
\label{1maKo}
Z(t|N) \sim
Z_{GKM}\left(L\left|\frac{1}{2}M^2 + N\log M\right.\right) \sim
\int_{m\times m} dM (\det M)^N
\exp \left[{\rm tr} \left(\frac{1}{2}M^2 + LM\right)\right]
\end{eqnarray}
In this representation analytic continuation to non-integer
values of $N$ is even more straightforward. Again, linear
combinations of solutions with different values of $N$, are
still solutions to the Virasoro constraints (\ref{virco}),
and all of them should be considered as particular branches
of the 1-matrix model partition function. This provides additional
information about the problem of conjugate variables:
$N$ itself can serve as one of them, or the coefficients of
above-mentioned linear combinations can -- changing the former
for the latter is a kind of
Fourier transform in the space of conjugate variables.

Possibility to represent one and the same partition function
by members of two different matrix model families, by (\ref{mamo})
and by (\ref{1maKo}), is a typical example of {\it duality}
between different families of quantum field theories.
One more duality follows from existence of CFT representations
for the same partition function, it relates these matrix models to
$2d$ free scalar theory and -- through one more duality -- to free
fermions. These dualities illustrate the fact that the nature and
the number of integration variables can differ in the most
radical way, still partition functions coincide: in absolutely
different theories one can find families of identical correlators.
Once again, this is an illustration of the general stringy idea:
{\it everything is the same}, one can fully describe one theory
(phase) in terms of another. Matrix models provide a nice
framework for testing this principle and can help to transform it
into constructive and reliable procedure.

\subsubsection{Limiting procedures and asymptotics}

Different realizations of partition functions are adequate or
at least convenient for descriptions of different phases of the
system (branches of partition function), or at least different
{\it classes} of phases. As usual, phases become pronouncely
different in particular asymptotics. Systematic approach implies
that all kinds of asymptotics should be investigated.
In particular, in the case of matrix models, all kinds of
continuum limits should be examined,
not only naive or t'Hooft's or double-scaling. Obviously, all
kinds of multi-cut solutions and all kinds of multi-scaling limits
require attention, but this is only the beginning: by no means
all possible asymptotics are exhausted in this way, and also there
is a lot of interesting beyond continuum limits. Unfortunately,
not too much is known about this variety of limits and nothing
like classification of asymptotics is available
(this was a trivial thing in the case of one variable and becomes
a pretty sophisticated issue in the case of infinitely many
variables).

There are two different possibilities to perform limiting
procedures: at the level of defining equations (linear
Ward identities or bilinear Hirota-style equations) and at the
level of particular realizations. Of course, one and the same limit
can look very different in different, though equivalent,
representations. Because the subject is under-investigated,
the set of known realizations is poor and many potentially
interesting asymptotics either do not attract attention or are
difficult to handle. We mention just two old problems, which still
remain unresolved.

In the class of generalized Kontsevich models (\ref{GKM}) and
(\ref{GKMKM}) one naturally distinguishes \cite{uniin1,GKMMM}
between {\bf Kontsevich phase} -- the asymptotics of large $L$ --
and the {\bf characters phase} -- the asymptotics of small $L$.
In these phases the "time-variables"
$t_k = r_k + \frac{1}{k}{\rm tr} L^{-k/n}$
with $n = {\rm deg}\ W'(z)$
are respectively close to $r_k = \frac{n}{k(n-k)}{\rm res}
\left(W'(z)\right)^{1-k/n}dz$ and to infinity.
Kontsevich phase corresponds, at least through the relation
(\ref{1maKo}), to perturbative phase of the models like
(\ref{mamo}), while the characters phase is the strong coupling
limit of these models. In this sense (\ref{1maKo}) is an
example of $S$-duality, interchanging weak and strong coupling
phases. Actually, these phases are closely
related to the strong and week coupling limits of Yang-Mills
theory and thus to confinement problem. However, this problem
was never exhaustively analyzed even at the level of matrix models
-- despite the existence of Kazakov-Migdal-Kontsevich family
(\ref{GKMKM}), which -- in variance with realizations like
(\ref{mamo}) -- provides an effective tool for studying both
asymptotics and interpolate between them (for these models
Kontsevich phase is the WKB asymptotics, while the characters
phase is perturbative limit). A systematic analysis, like
suggested in \cite{amm1,Eynard,amm23} for (\ref{mamo})
continues to wait for its time for Kontsevich and for more general
unitary matrix models.

Another abounded problem would be just a small paragraph in this
systematic analysis, still it is quite important by itself.
This is the problem of "double-scaling" limit \cite{ds} of
(\ref{mamo}), which is one of the simplest non-naive large-$N$
asymptotics of $Z(t|N)$, when $N\rightarrow\infty$ together with certain
special adjustment of $t$-variables. For investigation of this
limit one can make use of various techniques, of which the
most important are two: taking limit of loop equations
(Virasoro constraints) (\ref{virco}) \cite{Mak} and exploiting
the identity (\ref{1maKo}) and taking limit within the
family of Kontsevich models \cite{KMMMToda,ufn3}. In these ways
one can argue that
\begin{equation}
\label{dslim}
\lim_{\rm d.s.} Z(t|N) \sim \lim_{\rm d.s.}
Z_{GKM}\left(L\left|\frac{1}{2}M^2+N\log M\right)\right. =
Z_{GKM}^2\left(\tilde L\left|
\frac{1}{3}M^3\right)\right.
\end{equation}
where at the l.h.s. $t_{2k+1} = \frac{1}{2k+1}
{\rm tr} L^{-2k-1} = 0$ and $t_{2k} = \frac{1}{2k}
{\rm tr} L^{-2k}$, and the quantity at the r.h.s. (expressed
through
$\tilde t_{2k+1} = \frac{1}{k+1/2}{\rm tr}\tilde L^{-k-1/2}$)
has a special name of {\bf Kontsevich
$\tau$-function} $\tau_K(\tilde t)$ and is an increasingly important
matrix-model and string-theory special function.
However, even reliable derivation of the principally important
result (\ref{dslim})
and exact relation between $t$, $L$ and $\tilde t$, $\tilde L$
is not available, nothing to say about
corrections to this formula (describing what happens when one
{\it approaches} the limit) or its generalizations. Of special
importance among such generalizations
are the limits, when $Z(t|N) \rightarrow \tau_K^{2n}(\tilde t)$
with $n>1$, because of their obvious relation to Givental-style
decomposition formulas \cite{Giv,amm1,Kostov2} and  their
potential role in applications. It goes without saying that various
non-perturbative asymptotics of $\tau_K$ should be investigated:
this is the next natural task after such program is put on
track for the basic special function $Z(t|N)$.

\subsection{Integrability}

Integrability is believed \cite{ufn2,ufn3} to be the pertinent
property of quantum partition functions, reflecting the fact
that they are results of integration, which erases most of initial
information (roughly speaking, {\it almost all} forms are exact),
leaving only cohomological variables.
Therefore integrability is intimately related \cite{didi} to
topological-\-cohomological-\-holographic-\-stringy
theories and all these kinds of ideas should be considered
together. However, despite many efforts, spent on investigation of
particular models, we are still far from formulating clear
concepts. Partly this is because examples are dictated more by
applications than by the internal logic of the theory, and they
provide somewhat chaotic flow of information. There are already
several things that we seem to know, but their complete list
and relative significance remain obscure.

\subsubsection{Hirota equations}

The first in the list are bilinear Hirota-style equations, which
establish contact with Lie algebra theory \cite{KhL}.
Bilinearity is usually related to the properties of
comultiplication and to the basic relation $\Delta(g) =
g\otimes g$ for group elements \cite{MV,GMS}. Connes-Kreimer
theory \cite{GMS}-\cite{Mal} implies, that quantum partition functions
respect these properties, even if they are defined through
Feynman diagrams, without explicit reference to functional
integrals, and identify the underlying algebraic structure
with (generalized) renormalization group.
As already mentioned, deeper understanding of relation between
integrability and renormalization groups remains among the main
open problems of the theory.

\subsubsection{Moduli space of solutions}

The second big problem is understanding, classification and
control over the freedom in solving Hirota equations.
We mentioned already, that these bilinear equations restrict
the dependence on coupling constants $t_k$, and the
freedom remains in the conjugate/dual variables, like
zero-modes, or boundary conditions, or choices of vacua, or
holographic data, or whatever else name and analogy one
prefers to use in connection with one's favorite application.
The problem is to find not just a name, but adequate language
to speak about this data. Suggestive example is provided
by original Hirota equations \cite{Hir},
associated with KP hierarchy:
there the freedom (or at least a part of it) can be interpreted
in terms of Riemann surfaces. Different KP $\tau$-functions,
at least from the class of the {\it finite-zone solutions},
differ by the choice of Krichever data $\Sigma$ \cite{Kridata}:
a complex curve, a point on it and holomorphic coordinates
in the vicinity of a point. Given this data, the KP
$\tau$-function is fixed to be
\begin{equation}
\label{Kritau}
\tau(t|\Sigma) = e^{tQt}\theta (\sum_k\vec B_kt_k|T_{ij}),
\end{equation}
where $T_{ij}$ is the period matrix of the Riemann surface,
$\vec B_k = \oint_{\vec {\cal B}} \Omega_k$ and $\Omega_k(\zeta) =
\left(\zeta^{-(k+1)} + O(1)\right)d\zeta$ are meromorphic
1-forms with vanishing $A$-periods.
If ansatz (\ref{Kritau}) is substituted into Hirota equations,
they become equivalent to Shottky condition for the period
matrix $T_{ij}$ \cite{Nohyp}, i.e. require $\theta(\vec z|T_{ij})$
to be Riemannian theta function, not just arbitrary Abelian one.
Rather straightforwardly, the simplest -- rational and solitonic --
KP $\tau$-functions can be treated as particular cases of
(\ref{Kritau}) with Riemann surface of genus zero. All remaining,
non-finite-zone KP $\tau$-functions are more sophisticated
and can be considered as
associated with infinite-genus Riemann surfaces, though exact
meaning is not yet ascribed to these words. Clearly, there are many
different kinds of "infinite-genus" $\tau$-functions. Even the
principal example of Kontsevich $\tau$-function $\tau_K(t)$,
which belongs to this infinite-genus family, is not well understood
and investigated.

Anyhow, example (\ref{Kritau}) remains the main reference point
in the theory of integrable systems. It underlies the belief that
the right language for description of dual/conjugate variables to
times (coupling constants) should be that of Hodge theory, different
$\tau$-functions should be parameterized by data, encoded in
{\bf spectral surfaces}, not obligatory 2-dimensional. Thus
integrability is believed to be a unification of group theory and
algebraic geometry, with certain combinatorial flavor, already
coming from the studies of adjacent and clearly related subjects
\cite{DM,GMS,CK}. At the same time, even despite these ambitious
perspectives, too much remains obscure in the general structure
of integrability theory, see \cite{martmm}-\cite{dell}
for various important constituents which it should finally unify.
All this makes natural the fastly increasing
attention to integrability concepts, but decisive conceptual
breakthrough is still to come.

\subsubsection{Seiberg-Witten theory, Whitham integrability
and WDVV equations}

The third problem is the search for an adequate language, which
can help to merge Hodge theory with Lie algebra structures.
There are various approaches. The most obvious is to study
cohomological (topological) field and string theories \cite{tothe}
and find traces of integrability
there. This program is very successful in dealing with various
examples, but general concepts appear very hard to extract from it.
Among such concepts definitely are the WDVV equations
\cite{WDVV,WDVVMMM} and Batalin-Vilkovisky formalism \cite{BV,BVmod}.
Alternative approach,
from integrability side usually starts from (\ref{Kritau}),
directly or implicitly, and notes that theta-function in that
formula is oscillating, therefore one can distinguish between
fast (oscillating) and slow variables. Then, as usual, one can
take an average over fast variables and consider an effective theory
induced on the space of slow variables. Since, according to
(\ref{Kritau}), the fast variables are exactly the times $t_k$, the
slow variables of the emerging effective theory
should be exactly the dual variables, that
we are interested in. This idea, known as Whitham theory, has
obvious contacts with those of renormalization and
renormalization group, and not-surprisingly appears to have
immediate applications.  While absolutely un-developed in general,
it has a solid very well formulated chapter: the
Krichever-\-Hitchin-\-Seiberg-\-Witten theory, describing
explicit construction of peculiar special functions --
{\bf Seiberg-Witten prepotentials} -- from the analogue of Krichever
data, associated with peculiar Seiberg-Witten families of
spectral surfaces (the best known examples are Calabi-Yau spaces and
special families of Riemann surfaces). The prepotential
${\cal F}(S_k)$ is defined \cite{SW}-\cite{Ito} by two equations:
\begin{eqnarray}
\label{SWprep}
S_k = \oint_{{\cal A}_k} \Omega_{SW}, \nonumber \\
\frac{\partial{\cal F}}{\partial S_k} =
\oint_{{\cal B}_k} \Omega_{SW},
\end{eqnarray}
where $\Omega_{SW}(z)$ is a linear combination of $\Omega_k(z)$
in (\ref{Kritau}) and holomorphic 1-differentials $\vec\omega(z)$
(with non-vanishing $A$-periods), possessing a {\it special-geometry}
property that $\delta\Omega_{SW}$ is a symplectic form on the
"sheaf" of spectral surfaces over the moduli space. In practice
this condition states that $\delta\Omega_{SW}$ is holomorphic
(has no poles) if the variation is taken along the moduli
space of Seiberg-Witten curves.
Canonical system of contours $\{{\cal A}_k,{\cal B}_k\}$ in
(\ref{SWprep}) consists of
all non-contractable cycles (including those around
resolved singularities of $\Omega_k$) and their duals, see
\cite{Ito} for details. The variables $S_i$ define a sort of
{\it flat coordinates} on the moduli space.

It appears that the so defined
prepotential always satisfies the (appropriately generalized)
WDVV equations \cite{WDVVMMM}, namely for matrices
$\left(\hat{\cal F}_i\right)_{jk} = \frac{\partial^3{\cal F}}
{\partial S_i\partial S_j\partial S_k}$ made out
of prepotential's third derivatives,
\begin{equation}
\label{WDVV}
\hat{\cal F}_i\hat{\cal F}_j^{-1}\hat{\cal F}_k =
\hat{\cal F}_k\hat{\cal F}_j^{-1}\hat{\cal F}_i
\end{equation}
for any triple $(i,j,k)$. Moreover, this system of equations
for infinitely-large matrices often survives reduction
to finite-dimensional matrices when only $S$-variables,
associated with ordinary non-contractable contours, are
taken into account. This reasons why such reductions
exist are far from obvious,
and so is the entire theory of WDVV equations.
Technically, it relies upon {\it residue formulas} for
moduli derivatives of period matrices, but conceptually it
involves reduction theory of non-trivial {\it algebra of forms}
and is still very unsatisfactory.
Additional puzzle is that {\it reduced} WDVV equations are violated,
at least when applied naively, in the case of {\it elliptic}
Calogero system (see the third paper in ref.\cite{WDVVMMM}).
Other elliptic examples \cite{Skl,dell} were not yet analyzed in
the context of WDVV equations.

At the same time the theory of WDVV equations can appear even
more important than it seems. Today this set of equations is the
only candidate for a definition of a prepotential
in internal terms,
which does not refer to explicit construction procedure. Thus
WDVV equations can probably be promoted to the status of
{\it definition} of the prepotential and acquire the status,
similar to the one that Hirota equations have for $\tau$-functions.
(Referring to its possible origin as a Whitham average of
(\ref{Kritau}) -- or, more, accurately, of the underlying
Hirota equations,-- prepotential is often called
{\it Whitham or quasiclassical $\tau$-function}, and the
above-mentioned problem is to find the {\it internal} definition
of quasiclassical integrability -- a very important theoretical
challenge.)
The main problem with WDVV equations from this point of view
is that they describe only "spherical prepotentials":
from the study of topological theories and related algebraic
constructions \cite{Nek} we know that there is a whole
hierarchy of prepotentials, associated with the {\bf genus
expansion} of Yang-Mills and matrix models partition functions,
and Seiberg-Witten prepotential is just the zeroth (genus-zero)
term of this sequence. Appropriate generalization of WDVV
equations for entire sequence is not yet found, for first steps
in the cases of genus one and two see \cite{Getz}, and for
relation to Batalin-Vilkovisky theory see \cite{LoSh} (to avoid
possible confusion, these papers consider only WDVV equations
with additional requirement of "unit metric", which is natural
in applications to quantum cohomologies \cite{quco}
and naive topological models -- to "geometric
prepotentials",-- but is not quite straightforward to relax).

\subsection{CFT representations, Wick theorem and
decomposition formulas}

The theory of $2d$ conformal systems \cite{BPZ},
once at the front-line of attention in string theory,
was suddenly abandoned before an exhaustive theory was formulated,
and many problems, including effective description of entire
variety of conformal models and equivalencies (dualities) between
them, are left unresolved. Partly, this happened because attempts
to glue the entire subject together -- in the framework of
landscape theory \cite{GLM} -- naturally caused a desire
to interpolate between different conformal models, and this
unavoidably embeds \cite{cth} the world of $2d$ conformal theories
into that of $2d$ integrable systems -- still a very badly understood
subject, intimately related to the theory of quantum Kac-Moody
algebras \cite{Kh,BLZ,GKhL},
which attracts much less attention than it deserves.

As already mentioned, there is no doubt that every class of
matrix models have CFT representations, i.e. partition function
of a given family of matrix models can be represented as
a correlator of some operator in $2d$ conformal theory.
The questions are: what is the way to build this operator --
so far it is matter of art rather than a systematic procedure;--
what kind of information from the much bigger phase space
of $2d$ free fields is lost (projected out) in transition to
the matrix models (in particular, even models which are not
dual themselves can still produce representations of one and
the same matrix model partition function); how the data of
matrix model is mapped into the data of conformal model etc.
Today we do not know anything similar to answers to any of these
questions.

The crucial property of free field (and thus of $2d$ conformal)
theories is straightforward realization of the Wick theorem:
all correlators are decomposed into multi-linear combinations of
pair correlators (see \cite{bmm} for a rather fresh discussion
of the issue). In principle, for topological and integrable
theories naive Wick theorem is not applicable: even after
all possible simplifications there are contact terms
(and prepotential ${\cal F}(S)$ is almost never quadratic).
In interacting theory transformation to angle-action variables
which would explain to what kind of correlators the naive
Wick theorem is applicable, is highly non-linear and probably
not very useful. A very important question is what substitutes the
naive Wick theorem for generic partition functions. As we know,
generic partition functions are not at all in general position in
the space of all functions, they satisfy many constraints,
like linear Ward identities, bilinear Hirota equations, highly
non-linear WDVV equations. Thus we can expect that above question
about Wick theorem can make sense. Of course, one can try
to interpret as avatars of the Wick theorem all of above-mentioned
relations on partition functions. Remarkably, there is also
a much more profound candidate: decomposition formulas.
It appears that partition functions can often be represented
as poly-linear combinations of simpler building blocks, like
Gaussian matrix models and Kontsevich $\tau$-functions $\tau_K$.
So far this fact, which we call Givental-style decomposition,
was explicitly formulated in restricted number of cases
\cite{Giv,amm1,Kostov2}, but we believe that this is just the
tip of the iceberg, and further work will confirm the universality
and important role of decomposition phenomenon: a probable substitute
of Wick theorem for generic partition functions.

\section{Towards exhaustive theory of 1-matrix model:
the Dijkgraaf-Vafa theory}

Dijkgraaf-Vafa theory \cite{CIV}-\cite{Chekhov} deserves special
attention, because for the first time an application was found,
which requires understanding of a whole variety of phases of a
single matrix model, namely the model (\ref{mamo}). Despite
it is still not the {\it entire} variety -- only peculiar
phases are considered interesting, where partition function is
consistent with the {\bf genus expansion},  --
it is still a very big step forward, and it
stimulated fast progress in matrix model theory. In this
section we briefly characterize some directions of this progress
As everywhere in these notes, we concentrate on pure theoretical issues
and ignore the -- sometime very interesting -- results in applications
to Yang-Mills, quantum gravity and model building.

\subsection{Genus expansion}

Genus expansion, i.e. the t'Hooft's $1/N$ expansion \cite{thge},
attracts constant attention since the early studies of matrix
models \cite{loope}. While it has clear meanings both in diagram
technique (genus expansion for fat graphs) and in WKB approach
to the integrals like (\ref{mamo}), specification of partition
functions possessing genus expansion among generic solutions
of Ward identities \cite{virco} is not straightforward and looks
somewhat artificial. As explained in \cite{amm1,amm23}, it involves
several steps.

First of all, by rescaling of all time-variables $t_k \rightarrow
\frac{1}{g}t_k$ one introduces the new parameter $g$. It appears
in Virasoro constraints (\ref{virco}) as a coefficient $g^2$
in front of the double-derivative terms.
Second, solution of these Ward identities is looked for among
rather special functions, such that
\begin{equation}
g^2\log Z(t/g|N) = \sum_{p\geq 0} g^{2p}F^{(p)}(t|gN)
\end{equation}
is a series in non-negative powers of $g^2$, provided the
coefficients $F^{(p)}(t|gN)$ of this expansion depend on the t'Hooft's
coupling constant $S=gN$. Though consistent with
(\ref{virco}), this is actually a tricky requirement.
It implies that $Z$ itself is a series in all integer, positive
and negative, powers of $g^2$, just coefficients in front of
negative powers are strongly correlated. Moreover, there is no $N$
in Ward identities (\ref{virco}). $N$ can be introduced as
$S = gN = \partial (g^2\log Z)/\partial t_0$, but this requirement
breaks superposition principle: once it is imposed, the sum of two
solutions is not a solution (this is nearly obvious: a sum of two
exponentials is not an exponential). This means that genus expansion
can be at best a property of the elements of a linear {\it basis}
in the space of solutions, while generic element of this space does
not respect it. In other words, genus expansion can be a property
of particular integrals, like (\ref{mamo}), but there is no way
to impose this requirement on linear combinations of integrals
with different $N$, which will still be solutions of the Virasoro
constraints.

There are at least two ways to deal with this problem (the obvious
possibility to ignore it, as is usually done, is not counted).
First, one can look for the ways to define special basises --
the problem is to define them in invariant way, by specifying their
properties, not by explicit construction, like matrix integrals.
Second, one can try to formulate genus expansion without explicit
reference to $N$ -- this road leads to introduction of
{\bf check operators} \cite{amm23}.

\subsection{Gaussian and non-Gaussian partition functions}

Requirement of genus expansion is not enough to specify the
phase/branch of partition function completely, it regulates only
dependence on the common scaling factor of all couplings $t_k$ and
says nothing about their ratios. The next step \cite{amm1} is
to select combinations of $t$-variables which can appear in
denominators. A way to do this is to make a shift, $t_k \rightarrow
t_k-T_k$, and consider $W(\phi) = \sum_k T_k\phi^k$ as a {\bf bare
action} and $t_k\phi^k$ as perturbations. In other words, the
branch
\begin{equation}
\label{mamoW}
Z_W(t|N) = \frac{1}{{\rm Vol}(U(N))}
\int_{N\times N} d\Phi \exp \left(-{\rm Tr}W(\Phi) +
\sum_k t_k{\rm Tr}\ \Phi^k\right)
\end{equation}
of partition function is defined as a formal series
in non-negative powers of $t_k$, while $W(z)$, i.e. $T_k$'s, are
allowed to appear in denominator. Once $W(z)$ is introduced, one
immediately distinguishes between the Gaussian
(with $W(\phi) = \frac{M}{2}\phi^2$) and non-Gaussian phases.

Next, it turns out that while the variable $N$ is naturally
introduced in the Gaussian phase, in non-Gaussian case the situation
is more sophisticated: there are naturally $n = {\rm deg}\ W'(z)$
parameters like $N$: the phase is still split into a $n$-parametric
family of phases. Additional variables can be interpreted as
numbers of eigenvalues concentrated near the $n$ extrema (maxima
and minima) of the bare potential $W(\phi)$, a more reliable
interpretation is as integration constants in solution of shifted
(by $W(z)$) Virasoro constraints. See \cite{amm1} for further
details.

Gaussian partition function can be represented as
\begin{equation}
\label{Gapf}
Z_G^{(M)}(t|S|g) = \exp \left(\frac{1}{g^2}\left(-ST_0 -
\frac{M}{2}S^2\right)\right)
\exp\left(\sum_{p\geq 0} g^{2p-2}{\cal F}_G^{(p)}(t|S)\right)
\end{equation}
and information about the $t$-dependence of
prepotentials ${\cal F}_G^{(p)}(t|S)$ can be represented in
terms of {\bf multi-densities}, the generating functions of
$m$-point correlators, which -- for given $p$ and $m$ -- are
ordinary poly-differentials on the Riemann sphere.

Multi-densities can be defined in different ways (for
different families of correlators) \cite{amm1},
one of the ways consistent
with genus expansion and with the free-field representation
(\ref{virco}) of Virasoro constraints makes use of the
operator $\hat\nabla(z) = 2d\hat\phi_-(z) = \sum_{k=0}^\infty
\frac{dz}{z^{k+1}}\frac{\partial}{\partial t_k}$:
\begin{eqnarray}
\label{multdef}
\left.\left[\hat\nabla(z_1)\ldots\hat\nabla(z_m)
\left(g^2\log Z_W(t|\vec S|g)\right)\right]\right|_{t=0} =
\rho^{(\cdot|m)}(z_1,\ldots,z_m|\vec S|g) =
\sum_{p\geq 0} g^{2p}\rho^{(p|m)}(z_1,\ldots,z_m|\vec S)
\end{eqnarray}
The first {\it Gaussian} multi-densities are:
\begin{eqnarray}
\label{gamu}
\rho^{(0|1)}_G(z) = \frac{z - y_G(z)}{2},
\nonumber \\
\rho^{(1|1)}_G(z) = \frac{\nu}{y_G^5} =
-\frac{y_G^{\prime\prime}}{4y_G^2},
\nonumber \\
\rho^{(2|1)}_G(z) =
 \frac{5}{16}\frac{(y_G^{\prime\prime})^2}{y_G^5} -
\frac{1}{8y_G^2}\partial^2
\left(\frac{y_G^{\prime\prime}}{y_G^2}\right) -
\frac{1}{8}\frac{y_G^{\prime\prime\prime\prime}}{y_G^4},
\nonumber \\ \nonumber \\
\ldots
\end{eqnarray}
\begin{eqnarray}
\rho^{(0|2)}_G(z_1,z_2) = \frac{1}{2(z_1-z_2)^2}
\left(\frac{z_1z_2-4\nu}{y_G(z_1)y_G(z_2)} -1\right) =
\nonumber \\
 = -\frac{1}{2y_G(z_1)}
\frac{\partial}{\partial z_2}\frac{y_G(z_1) - y_G(z_2)}{z_1-z_2}
 = -\frac{1}{2y_G(z_2)}
\frac{\partial}{\partial z_1}\frac{y_G(z_1) - y_G(z_2)}{z_1-z_2},
\nonumber
\end{eqnarray}
\begin{eqnarray}
\rho^{(1|2)}_G(z_1,z_2) =
\frac{\nu}{y_G^7(z_1)y_G^7(z_2)}
\Big(
z_1z_2(5z_1^4 + 4z_1^3z_2 + 3z_1^2z_2^2 + 4z_1z_2^3 + 5z_2^4) +
 \nonumber \\  +
4\nu\left[z_1^4 - 13z_1z_2(z_1^2 + z_1z_2 + z_2^2) + z_2^4\right] +
16\nu^2(-z_1^2 + 13z_1z_2 - z_2^2) + 320\nu^3
\Big)=\nonumber\\
=\frac{1}{ y_{G1}}\left[
\left(4\frac{1}{4 y_{G1}^2} y_{G1}''
-\frac{1}{2 y_{G1}}\partial_1^2\right)
\frac{1}{2 y_{G1}}\frac{\partial}{\partial z_2}\frac{ y_{G1}-
y_{G2}}{(z_1-z_2)}+\right.\nonumber\\
+\left.\frac{\partial }{\partial z_2}\frac{1}{z_1-z_2}
\left(\frac{1}{4 y_{G2}^2} y_{G2}''-\frac{1}{4 y_{G1}^2} y_{G1}'' + \ \
\frac{1}{y_{G1}}\left(-\frac{1}{4 y_{G1}} y''_{G1}+
\frac{1}{2 y_{G1}}\frac{\partial}{\partial z_2}\frac{ y_{G1}-
y_{G2}}{(z_1-z_2)}\right)
\right)\right],
\nonumber \\ \nonumber \\
\ldots
\nonumber
\end{eqnarray}
\begin{eqnarray}
\rho^{(0|3)}_G(z_1,z_2,z_3) =
\frac{2\nu (z_1z_2 + z_2z_3 + z_3z_1 + 4\nu)
}{y_G^3(z_1)y_G^3(z_2)y_G^3(z_3)}
= \nonumber \\
=\frac{1}{y_{G1}}\left[
2\frac{1}{4y_{G1}^2}\left(\frac{\partial}{\partial z_2}
\frac{y_{G1} -  y_{G2}}{z_1-z_2}
\right)\left(\frac{\partial}{\partial z_2}
\frac{y_{G1} -  y_{G3}}{z_1-z_3}\right)+\right.\nonumber\\
+\frac{\partial}{\partial z_2}\frac{1}{z_2-z_1}\left(
\frac{1}{2y_{G1}}\left(\frac{\partial}{\partial z_3}
\frac{y_{G1} -  y_{G3}}{z_1-z_3}
\right)-
\frac{1}{2y_{G2}}\left(\frac{\partial}{\partial z_3}
\frac{y_{G2} -  y_{G3}}{z_2-z_3}\right)
\right)+\nonumber\\
+\left.\frac{\partial}{\partial z_3}\frac{1}{z_3-z_1}\left(
\frac{1}{2y_{G1}}\left(\frac{\partial}{\partial z_2}
\frac{y_{G1} -  y_{G2}}{z_1-z_2}
\right)-
\frac{1}{2y_{G3}}\left(\frac{\partial}{\partial z_2}
\frac{y_{G2} -  y_{G3}}{z_2-z_3}\right)
\right) \right],
\nonumber \\ \nonumber \\
\ldots \nonumber
\end{eqnarray}
We assumed here that $M=1$, then $y^2_G = z^2-4\nu$ and
$\nu = S = gN$.

Non-Gaussian are somewhat more sophisticated,
they are actually made from Riemann theta-functions for
a peculiar Seiberg-Witten family of
{\bf hyperelliptic complex curves}. For example, the $(p,m) =
(0,2)$-density is the $2$-point correlator on such
surfaces:
\begin{eqnarray}
\label{noga2}
\rho_W^{(0|2)}(z_1,z_2) = d_{z_1}d_{z_2}\log E(z_1,z_2),
\end{eqnarray}
where $E(z_1,z_2) \sim \frac{\nu_*(z_1)\nu_*(z_2)}
{\theta(\vec z_1-\vec z_2)}$
is the prime form \cite{Fay,MP}.

\subsection{Decomposition formulas}

Rewriting (\ref{mamoW}) in terms of the eigenvalues and then
separating the eigenvalues into sets, associated with different
extrema of $W(\phi)$, one can deduce \cite{KMT} a decomposition
formula \cite{amm1} for non-Gaussian $Z_W$:
\begin{eqnarray}
\label{deco}
Z_W  \sim
\frac{\prod_{i=1}^n e^{-N_iW(\alpha_i)}{\rm Vol}(U(N_i))}
{{\rm Vol}(U(N))}
\prod_{i<j}^n\alpha_{ij}^{2N_iN_j}\hat O_{ij}
\prod_{i=1}^n \hat O_i\ \prod_{i=1}^n
Z_G^{(M_i)}(t^{(i)}|N_i)
\end{eqnarray}
with $\alpha_i$ denoting the $n$ roots of
$$
W'(z) = \sum_{k=0}^{n+1}kT_kz^{k-1} =
(n+1)T_{n+1}\prod_{i=1}^n (z-\alpha_i),
$$
$\alpha_{ij} = \alpha_i - \alpha_j$, the frequencies
$\M_i = W''(\alpha_i) =
(n+1)T_{n+1}\prod_{j\neq i}\alpha_{ij}$, and operators
$$
\hat{\cal O}_{ij} =
\exp \left(2\sum_{k,l=0}^\infty (-)^{k}\frac{(k+l-1)!}
{\alpha_{ij}^{k+l}k!l!}
\frac{\partial}{\partial t^{(i)}_k}
\frac{\partial}{\partial t^{(j)}_l}
\right),
$$ $$
\hat{\cal O}_i = \exp \left(-\sum_{k\geq 3}
\frac{\partial^kW(\alpha_i)}{k!}
\frac{\partial}{\partial t^{(i)}_k}\right)
$$
As explained in the previous section, this
Givental-style formula is an example
of general phenomenon, generalizing the Wick theorem from
free fields and conformal theories to matrix models and
non-trivial integrable systems.

Eq.(\ref{deco}) requires (and deserves) deep investigation.
Today almost nothing is known about it. It is unclear how (\ref{deco})
follows directly from Virasoro constraints, without explicit
use of matrix integrals. It is unclear what is exact implication
of the relation (\ref{dslim}) for Gaussian functions,
standing at the r.h.s. of (\ref{deco}). It is clear that
the combination of (\ref{dslim}) and (\ref{deco}) should
represent $Z_W$ as some operator acting on $\tau_K^{2n}$, but
exact formula is unavailable (Greg Moore in \cite{Moore}
and especially Ivan Kostov in \cite{Kostov2} came very close
to the answer, but it still escapes).

Eq.(\ref{deco}) can be also considered as a definition of
a linear basis in the space of solutions of Virasoro constraints,
which is formed by the functions, respecting the genus expansion.
{\it Arbitrary} solution can be represented as a linear combination,
\begin{equation}
\label{genso}
Z_W(t) = \int_{\vec S} \mu(\vec S)Z_W(t|\vec S),
\end{equation}
with {\it arbitrary} measure $\mu(\vec S)$.

\subsection{Check-operators and $t$-evolution operator}

According to their definition (\ref{multdef}), multi-densities
do not depend on the time variables $t_k$, only on $T_0,\ldots,
T_{n+1}$ and $\vec S$. Actually, dependence on $T_n$ and $T_{n+1}$
is fixed by (\ref{virco}), and multi-densities depend on $2n$
variables: $n$ of them are $T$'s and $n$ are $S$'s.
Therefore one can ask for a procedure which defines multi-densities
directly in terms of these variables, without an intermediate
introduction of -- infinitely many -- auxiliary (from the point
of view of multi-densities theory) variables $t_k$. This
problem is solved, at least conceptually, in
\cite{amm1,amm23} in terms of {\bf check-operators}.

Multi-densities are not arbitrary functions of their $2n$ variables,
the Ward  identities (\ref{virco}) express all of them in terms
of a single function of $n+1$ variables, $T_0,\ldots,T_{n-1}$ and $g$
\cite{amm1}, which we call {\bf bare prepotential}. The choice
of bare prepotential is actually the choice of particular branch
of partition function. Note, that this means that there are not just
infinitely many branches, they form a continuous variety -- and this
is a property of every reasonable partition function and of anything
obtained by application of {\bf evolution operators} (for which
the functional integral is a particular realization),
see \cite{DM} for explanation, how continuous phase structure emerges
from discrete bifurcations (phase transitions), and for discussion
of the adequate formalism. This set of questions is very important,
in particular it can help to understand the {\it relevant} topology
and may be even the metric structure on the space of phases, what is
the principal problem in landscape theory.
Development of such formalism for
matrix models is a task for the future, today we know how to
proceed within a given phase, when bare prepotential is somehow
specified and the question of relative importance of different
choices is not addressed.

\subsubsection{Independent variables and bare prepotential}

To understand what it means to specify the bare prepotential,
we consider immediate corollary
of (\ref{virco}) with $t_k \rightarrow \frac{1}{g}(t_k-T_k)$
for partition function at vanishing times $t_k$.
It is given by \cite{amm23}:
\begin{equation}
\label{etaparam}
\left.Z(T|g)\right|_{t=0} = \int dk z(k|\eta_2,\ldots,\eta_n|g^2)
e^{\frac{1}{g^2}(kx-k^2w)}
\end{equation}
with an arbitrary function $z$ of $n$ arguments $(k,\eta_2,\ldots,\eta_n)$
and $g^2$.
Here the $\hat L_{-1}$-invariant variables are used,
\begin{equation}
w = \frac{1}{n+1}\log T_{n+1}, \ \ \ \
\eta_k = T_{n+1}^{-\frac{nk}{n+1}} \left( T_n^k + \ldots \right),\ \ \ \
x  = T_0 + \ldots \sim \eta_{n+1}
\end{equation}
In particular,
$$\eta_2  = \left(T_n^2 -
\frac{2(n+1)}{n}T_{n-1}T_{n+1}\right)T_{n+1}^{-\frac{2n}{n+1}},
$$ $$
\eta_3  = \left(T_n^3 -
\frac{3(n+1)}{n}T_{n-1}T_nT_{n+1} +
\frac{3(n+1)^2}{n(n-1)}T_{n-2}T_{n+1}^2
\right)T_{n+1}^{-\frac{3n}{n+1}},
$$ $$
\eta_4  = \left(T_n^4 -
\frac{4(n+1)}{n}T_{n-1}T_n^2T_{n+1} +
\frac{8(n+1)^2}{n(n-1)}T_{n-2}T_nT_{n+1}^2 -
\frac{8(n+1)^3}{n(n-1)(n-2)}T_{n-3}T_{n+1}^3
\right)T_{n+1}^{-\frac{4n}{n+1}},
$$ $$
\ldots $$
$$
\eta_k = \left(T_n^k + \frac{k(k-2)!}{n!}\sum_{l=1}^{k-1} (-)^l
\frac{(n+1)^l (n-l)!}{(k-l-1)!}
T_{n-l}T_n^{k-l-1}T_{n+1}^l
\right)T_{n+1}^{-\frac{kn}{n+1}}
$$
The variable $x$ is obtained from $\eta_{n+1}$ by normalization
and it is the only variable which contains $T_0$.
At the same time only $T_0$ appears in double-derivative
item of $\hat L_0$, thus in (\ref{etaparam}) the
$\hat L_{0}$-constraint links the $x$- and $w$-dependencies.
Of course, generic (\ref{etaparam}) does not possess genus
expansion, i.e. does not guarantee that the bare prepotential
$F(T|g) = g^2\log Z(T|g)$, is expanded in non-negative powers
of $g^2$: as explained above, this requirement is somewhat
artificial from the perspective of $D$-module theory.
Still, a simple ansatz makes things consistent (it is not
quite clear if this ansatz is absolutely necessary for
genus expansion to exist): if
\begin{equation}
S = \frac{\partial{ F}}{\partial T_0} = const
\end{equation}
i.e. is independent of $T_0,\ldots,T_{n+1}$ and $g$,
then,
\begin{equation}
z(k|\eta_2,\ldots,\eta_n|g^2) =
H(\eta_2,\ldots,\eta_n|S|g^2)\ \delta(k-S)
\end{equation}
and
\begin{equation}
F = g^2\log Z = Sx + \frac{S^2}{n+1}\log T_{n+1} + g^2\log
H(\eta_2,\ldots,\eta_n|S|g^2)
\end{equation}
where $H$ is an arbitrary function of $n+1$ variables.
Genus expansion occurs, if we request $g^2\log H$ is expanded in
non-negative powers of $g^2$.

Introduction of the other $\vec S$-variables
can be considered as ingenious version of Fourier-Radon transform
with the help of Dijkgraff-Vafa partition
functions (\ref{deco}), which converts $H(\eta_2,\ldots,\eta_n)$ into
an arbitrary measure $\mu(\vec S)$ of the $\vec S$-variables
in (\ref{genso}) \cite{amm1}.

\subsubsection{Check-operator multi-densities}

Other $\hat L_k$-constraints with $k > 0$
express $t$-dependencies of $Z(t|T)$ through $T$-dependencies.
In particular, they allow to represent multi-densities (both
Gaussian and non-Gaussian) by action of $T$-dependent operators
on partition function (\ref{etaparam}). Such operators,
involving only derivatives with respect to $T$-variables, are
named {\it check-operators} in \cite{amm23}, to distinguish them
from {\it hat-operators}, acting on $t$-variables. The task
is to express multi-densities, defined by application of
hat-operators to the full prepotential $F(t|T|g)$,
through check-operators applied to the bare prepotential $F(T) =
F(t=0|T|g)$. These check operators will predictably be more
sophisticated than hat-operators $\hat\nabla(z)$, but instead
they can be applied to the {\it independent} (free) function
$F(T)$.

Remarkably, the problem of building check-operator multi-densities
appears essentially equivalent to the problem of {\it Gaussian
multi-densities} \cite{amm23}. Though this fact is not fully proved
-- and even adequately formulated -- yet, it is
not too surprising: the needed check-operators are universal,
in certain sense they
do not depend on the phase, thus they should be restorable from
information, available in any phase, including the Gaussian one
(once again we encounter the main idea of string theory).
Anyhow, modulo some details concerning ordering prescriptions,
the first check-operators can be just read from formulas
(\ref{gamu}) for Gaussian multi-densities \cite{amm23}:
\begin{eqnarray}
\label{comu}
\check\rho_W^{(0|1)}(z|g) = \frac{W'(z) - \check y(z|g)}{2},
\nonumber \\
\check\rho_W^{(1|1)}(z|g) =
-\frac{1}{4\check y^2}\check y^{\prime\prime},
\nonumber \\
\check\rho_W^{(2|1)}(z|g) =
\frac{5}{16}\frac{(\check y^{\prime\prime})^2}{\check y^5} -
\frac{1}{8\check y^2}
\partial^2\left(\frac{\check y^{\prime\prime}}{\check y^2}\right) -
\frac{1}{8}\frac{\check y^{\prime\prime\prime\prime}}{\check y^4},
\nonumber \\ \nonumber \\
\ldots
\nonumber \\ \nonumber \\
\check\rho_W^{(0|2)}(z_1,z_2|g) = -\frac{1}{2\check y(z_1|g)}
\frac{\partial}{\partial z_2}
\frac{\check y(z_1|g) - \check y(z_2|g)}{z_1-z_2},
\nonumber \\
\check\rho_W^{(1|2)}(z_1,z_2|g) =
\frac{1}{\check y_1}\left[
\left(4\frac{1}{4\check y_1^2}\check y_1''
-\frac{1}{2\check y_1}\partial_1^2\right)
\frac{1}{2\check y_1}\frac{\partial}
{\partial z_2}\frac{\check y_1-\check y_2}{(z_1-z_2)}
+\right.\nonumber\\
+\left.\frac{\partial }{\partial z_2}\frac{1}{z_1-z_2}
\left(\frac{1}{4\check y_2^2}\check y_2''-
\frac{1}{4\check y_1^2}\check y_1''+
\frac{1}{y_1}\left(-\frac{1}{4\check y_1}\check y''_1+
\frac{1}{2\check y_1}\frac{\partial}{\partial z_2}
\frac{\check y_1-\check y_2}{(z_1-z_2)}\right)
\right)\right],
\nonumber \\ \nonumber \\
\ldots
\nonumber\\ \nonumber \\
\check\rho_W^{(0|3)}(z_1,z_2,z_3|g) =
\frac{1}{\check y_1}\left(
2\frac{1}{2\check y_1}\left(\frac{\partial}{\partial z_2}
\frac{\check y_1 -  \check y_2}{z_1-z_2}
\right)\frac{1}{2\check y_1}\left(\frac{\partial}{\partial z_2}
\frac{\check y_1 -  \check y_3}{z_1-z_3}\right)+\right.\nonumber\\
+\frac{\partial}{\partial z_2}\frac{1}{z_2-z_1}\left(
\frac{1}{2\check y_1}\left(\frac{\partial}{\partial z_3}
\frac{\check y_1 -  \check y_3}{z_1-z_3}
\right)-
\frac{1}{2\check y_2}\left(\frac{\partial}{\partial z_3}
\frac{\check y_2 -  \check y_3}{z_2-z_3}\right)
\right)+\nonumber\\
+\left.\frac{\partial}{\partial z_3}\frac{1}{z_3-z_1}\left(
\frac{1}{2\check y_1}\left(\frac{\partial}{\partial z_2}
\frac{\check y_1 -  \check y_2}{z_1-z_2}
\right)-
\frac{1}{2\check y_3}\left(\frac{\partial}{\partial z_2}
\frac{\check y_2 -  \check y_3}{z_2-z_3}\right)
\right)\right),
\nonumber\\ \nonumber \\
\ldots
\end{eqnarray}
Here
\begin{equation}
\check y^2(z|g) = W'(z)^2 + 4g^2\check R_W(z)\ \ \ {\rm and} \ \ \
\check R_W(z) = \sum_{a,b=0} (a+b+2)T_{a+b+2}z^a
\frac{\partial}{\partial T_b}
\end{equation}
The check-multidensities are defined to satisfy
$$ \check K_W^{(\cdot|m)}(z_1,\ldots,z_m|g) Z_W(T|g) =
K_W^{(\cdot|m)}(z_1,\ldots,z_m|g) Z_W(T|g) =
$$ $$ =
\left.\left[\hat\nabla(z_1)\ldots \hat\nabla(z_m) Z_W(t|T|g)\right]
\right|_{t=0}
$$
for {\it full} correlators $K_W^{(\cdot|m)}(z_1,\ldots,z_m|g)$,
which are derivatives of $Z(t)$, not of its logarithm,
while for {\it connected} $\rho_W^{(\cdot|m)}(z_1,\ldots,z_m|g)$
$$\left(Z_W(T|g)\right)^{-1}
\check \rho_W^{(\cdot|m)}(z_1,\ldots,z_m|g)\ Z_W(T|g) \neq
\rho_W^{(\cdot|m)}(z_1,\ldots,z_m|g) =
$$ $$ =
\hat\nabla(z_1)\ldots \hat\nabla(z_m) \left(g^2\log Z_W(T|g)\right)
$$
Relation between the full and connected correlators is
provided by straightforward -- though heavily looking --
combinatorial formula \cite{amm23}
\begin{eqnarray}
K_W^{(\cdot|m)}(z_1,\ldots,z_m|g) =
\sum_\sigma^{m!} \ \sum_{k=1}^{m} \
\sum_{\nu_1,\ldots,\nu_k = 1}^\infty
\sum_{p_1,\ldots,p_\nu=0}^\infty g^{2(p_1+\ldots +p_\nu-\nu)}
\nonumber \\
\left(\
\sum_{\stackrel{m_1,\ldots,m_k}{m = \nu_1m_1+\ldots+\nu_km_k}}
\frac{1}{\nu_1!(m_1!)^{\nu_1}\ldots \nu_k!(m_k!)^{\nu_k}}
\rho_W^{(p_1|\tilde m_1)}(z_{\sigma(1)},\ldots,z_{\sigma(\tilde m_1)})
\cdot\right.\nonumber \\
\hspace{-1cm} \left.  \phantom{\sum^{5^{5^{5^{5^{5^5}}}}}}
\cdot \rho_W^{(p_2|\tilde m_2)}
(z_{\sigma(\tilde m_1+1)},\ldots,z_{\sigma(\tilde m_2)})\ldots
\rho_W^{(p_\nu|\tilde m_\nu)}
(z_{\sigma(m-\tilde m_\nu+1)},\ldots,z_{\sigma(m)})
\ \right),
\end{eqnarray}
which serves as generalization of the Wick formula from the
case when the only connected correlators are $2$-point
(note that this simplification does not occur even in Gaussian
phases of matrix models beyond naive continuum limit).
A similar expression relates the check-operators $\check K$
and $\check \rho$, in particular,
\begin{equation}
\check K_W^{(\cdot|1)}(z;g) =
\sum_{p=0}^\infty g^{2p-2}\check\rho_W^{(p|1)}(z;g)
\end{equation}
Note also, that in variance with (\ref{gamu}),
expressions in (\ref{comu}) explicitly contain $g^2$.

Actually, once the check-operators multi-densities
are introduced, one can forget about the genus-expansion
constraint: operators can be applied to any bare
partition function. Instead of being a constraint
on partition functions, genus expansion requirement
dictates becomes a selection criterion for a basis in
check-operators space. The next step is to build up a theory
of the {\bf $t$-evolution operator} $\check U(t)$, which
generates the $t$-dependence of partition function:
\begin{equation}
Z(t|T) = \check U(t)Z(0|T).
\end{equation}
Its existence is almost obvious, see \cite{amm1,amm23},
but properties and explicit realizations remain to be found.

\subsection{Seiberg-Witten theory}

The Dijkgraaf-Vafa theory is intimately related to
Seiberg-Witten theory.

\subsubsection{Genus-zero level}

For detailed discussion of this relation
in the particular case of Dijkgraaf-Vafa partition
functions {\it per se}, i.e. with requirements that
genus expansion exists and only
genus-zero contributions (spherical prepotentials)
are considered, see \cite{ItoM1}-\cite{CMMV}.
Here are the main points of this analysis.
The smallest moduli space of Seiberg-Witten structure
unifies all phases with all the bare potentials $W(z)$
of a fixed degree $n$. It has (complex) dimension $2n$,
$T$ and $S$-variables form the set of
$2n$ {\it flat moduli} of Seiberg-Witten structure,
defined with the help of $\Omega_{SW} = \sqrt{W'(z)^2 -
4f_W(z)}dz$, where $f_W(z) = -\check R_W(z)F^{(0)}(T)$
is a polynomial of degree $n-1$, made from arbitrary
function $F^{(0)}(T)$ of $n$ variables $T_0,\ldots,
T_{n-1}$ (and dependence on $T_{n}$ and $T_{n+1}$ prescribed
by $\check L_{-1}$ and $\check L_0$-constraints).
The space of all functions of $n$ variables can --
in the framework of Fourier-Radon transforms -- be
parameterized by arbitrary functions of $n$ other parameters,
and $\vec S$ can play the role of such parameters no better,
no worse than
any other set of variables. What distinguishes $\vec S$
is that they are flat Seiberg-Witten moduli, i.e. the
symplectic form on the sheaf of Seiberg-Witten curves is
\begin{equation}
\delta \Omega_{SW} = \omega_i(z) \wedge \delta S_i  +
\Omega_k(z) \wedge \delta T_k
\end{equation}
and
\begin{equation}
\vec S = \oint_{\vec{\cal A}}\Omega_{SW},\ \ \
T_k = {\rm res}_\infty z^k\Omega_{SW}
\end{equation}
The CIV-DV prepotential \cite{CIV,DV} is defined by
\begin{equation}
\label{CIVDV}
\frac{\partial{\cal F}_{CIV}}{\partial\vec S} =
\oint_{\vec{\cal B}}\Omega_{SW},\ \ \
\frac{\partial{\cal F}_{CIV}}{\partial T_k} =
\frac{1}{k}{\rm res}_\infty z^{-k}\Omega_{SW}
\end{equation}
it can be explicitly calculated term by term and possesses pronounced
group-theoretical structure \cite{ItoM3,amm1}, which still awaits its
adequate interpretation and explanation. No reasonable generic
formulas for this prepotential are yet found, despite it is a
genus-zero quantity, and despite it can be deduced in at least two dual
ways: from (\ref{CIVDV}) and from (\ref{deco}). Moreover, even
direct relation between these equivalent representations is not
yet established. Also important is that there are various differently
looking forms for the same items in ${\cal F}_{CIV}$, their equivalence
follows from peculiar identities \cite{ItoM4}, which are simple to
prove, but not so simple to understand.

\subsubsection{WDVV equations}

In \cite{ItoM1,ItoM4,CMMV} the proof is given
of the WDVV equations (\ref{WDVV}) for the CIV-DV prepotential.
This is an additional check that $2n$ moduli $T$ and $S$ form
complete set, at least at the level of genus-zero prepotentials.
In \cite{ItoM2} a less obvious
observation is reported:  sometime this set can be reduced further,
for example, by elimination of one of the $4$ moduli in the
case of $n=2$, however, the meaning of this observation remains
obscure.

The analogues of the CIV-DV prepotential for higher genera
can also be defined because of its relation to Ward identities
(\ref{virco}) -- this makes the situation different from
generic Seiberg-Witten theory, where the problem of lifting
from genus zero to all genera remains very important, but
unsolved. This makes it possible to ask, if higher genus
prepotentials satisfy the (properly generalized) WDVV equations,
for example the ones
implied by studies of topological models \cite{Getz}. This
problem was not seriously attacked yet.

\subsubsection{Towards quantum Seiberg-Witten theory of check
multi-densities}

The next natural question in the theory of check operators
concerns definition of $S$-variables. Since in particular phases
they are periods of $\rho_W^{(0|1)}$, it is natural to
treat $S$-variables as {\it operators}, defined as periods of
$\check\rho^{(0|1)}$. Moreover, under certain conditions the
$A$-periods of higher-genus one-point densities are vanishing,
and one can even consider the periods of
$\check\rho^{(\cdot|1)}$ or
\begin{equation}
\check S_i = \frac{g^2}{4\pi i}
\oint_{{\cal A}_i}\check K^{(\cdot|1)}(z).
\end{equation}
This is a very perspective direction of research. Among other,
obvious and not-so-obvious, questions is developement of the
full-scale version of Seiberg-Witten theory for check-operator
quantities, i.e. a quantization of Seiberg-Witten structure.
Introduction of check operators with the goal to deal with
arbitrary bare prepotentials in a uniform way is a nice illustration
of string "third-quantization" procedure, which implies that in
order to study {\it families} of models one actually needs to
quantize a representative of the family.
An example of the simplest relation
in this {\it quantum Seiberg-Witten theory} is provided by the
quantum version of (\ref{SWprep}) \cite{amm23}:
\begin{equation}
\label{quaco}
\left[\frac{g^2}{4\pi i}\oint_{{\cal A}_i}\check K,\
\frac{g^2}{2}\oint_{{\cal B}_j}\check K\right]
= g^2\delta_{ij},
\end{equation}
probably, with no corrections at the r.h.s.

The check-operator $A$-and $B$-periods
in (\ref{quaco}) are equal respectively to
$\check S_i = \check S_i^{(0)} + g^2\check S_i^{(1)} + O(g^4)$
and $\check \Pi_i^{(0)} + O(g^2)$,
where \cite{ItoM3}
\begin{equation}
\label{pers}
\check S_i^{(0)} = \frac{g^2}{\M_i}\check R_i,
\ \ \  \ \ \ \ \check\Pi_i^{(0)} = W(\alpha_i),
\end{equation}
$$
g^2\check S_i^{(1)}\ \stackrel{(\ref{comu})}{=}\ {\rm res}_{\alpha_i}
\frac{W^{\prime\prime\prime}(z)dz}{8W'(z)^2} =
\frac{\partial^4W \partial^2 W - (\partial^3W)^2}{8(\partial^2 W)^3}
(\alpha_i)
$$
and $\alpha_i$, $i = 1,\ldots,n$ are the roots of $W'(z)$,
operators $\check R_i = \check R_W(\alpha_i)$,
$$\M_i = W''(\alpha_i) =
(n+1)T_{n+1}\prod_{j\neq i}\alpha_{ij},$$
$\alpha_{ij} = \alpha_i - \alpha_j$.
Note that in the leading order the operator $B$-periods and
genus-$1$ contributions to $A$-periods
do not containt $T$-derivatives, thus non-trivial are only commutators,
involving $\check S_i^{(0)}$. These commutators can
be easily deduced from two general formulas from \cite{amm23},
\begin{equation}
\label{comr}
\left[\check R_W(x), W(z)\right] = \frac{W'(x)-W'(z)}{x-z}, \ \ \ \ \ \
\left[\check R_W(x), \check R_W(z)\right] =
(\partial_z-\partial_x)\frac{\check R_W'(x)-\check R_W'(z)}{x-z},
\end{equation}
together with
\begin{equation}
\frac{\partial\alpha_j}{\partial T_k} =
-\frac{k\alpha_j^{k-1}}{\M_j},
\end{equation}
from \cite{ItoM4}, which implies straightforwardly:
$$\left[\check R_W(x), \alpha_j \right] =
\sum_{a,b=0} (a+b+2)T_{a+b+2}x^a \frac{\partial\alpha_j}{\partial T_b} =
-\frac{1}{\M_j}\left.\partial_z \frac{W'(x) - W'(z)}{x-z}
\right|_{z=\alpha_j}$$
and, since $W'(\alpha_i) = W'(\alpha_j) = 0$,
\begin{equation}
\label{ral}
\left[\check R_i, \alpha_j \right] =
\left[\check R_W(\alpha_i), \alpha_j \right] = \frac{1}{\alpha_{ij}}
\end{equation}
Eqs.(\ref{comr}) need to be supplemented by (\ref{ral}), because
in commutators of periods (\ref{pers}) the arguments $\alpha_i$
also depend on $T_k$ and their derivatives should be taken into
account. Thus instead of (\ref{comr}) we need:
\begin{eqnarray}
\label{RM}
\left[\check R_i, \M_j\right] =
\left[\check R_W(\alpha_i), W''(\alpha_j)\right] =
\left.\partial_z^2\frac{W'(\alpha_i)-W'(z)}
{\alpha_i-z}\right|_{z=\alpha_j} +
W^{\prime\prime\prime}(\alpha_j)
\left[\check R_i, \alpha_j \right] =
-\frac{2M_j}{\alpha_{ij}^2},
\end{eqnarray}
and
\begin{eqnarray}
\left[\check R_i, \check R_j\right] =
\left[\check R_W(\alpha_i), \check R_W(\alpha_j)\right] =
\left.(\partial_z-\partial_x)\frac{\check R_W(x)-\check R_W(z)}{x-z}
\right|_{x=\alpha_i,\ z = \alpha_j} +
\nonumber \\ +
\left[\check R_W(\alpha_i), \alpha_j \right]\check R_W'(\alpha_j) -
\left[\check R_W(\alpha_j), \alpha_i \right]\check R_W'(\alpha_i) =
\frac{2}{\alpha_{ij}^2}
\left(\check R_i - \check R_j\right)
\end{eqnarray}
Now we have everything to check, that
$$ \left[\check S_i^{(0)}, \check S_j^{(0)}\right] =
\left[\frac{g^2}{\M_i}\check R_i, \frac{g^2}{\M_j}\check R_j\right]
= \frac{g^4}{\M_i\M_j}\left(\left[\check R_i, \check R_j\right] -
\frac{1}{\M_j}\left[\check R_i, \M_j\right]\check R_j +
\frac{1}{\M_i}\left[\check R_j, \M_i\right]\check R_i \right) = 0$$
Direct generalization of (\ref{RM}) states, that
$$
\left[\check R_i, \partial^mW(\alpha_j)\right] = -m!
\sum_{k=0}^{m-2}\frac{1}{(k+1)!}\frac{\partial^{k+2}W(\alpha_j)}
{\alpha_{ij}^{m-k}},
$$
in particular,
$$
\left[\check R_i,
\frac{\partial^4W \partial^2 W - (\partial^3W)^2}{(\partial^2 W)^3}
(\alpha_j)\right] = -\frac{24}{M_j\alpha_{ij}^2},
$$
so that
$$
\left[\check S_i^{(0)}, g^2\check S_j^{(1)}\right] =
-\frac{3}{M_iM_j\alpha_{ij}^2} =
\left[\check S_j^{(0)}, g^2\check S_i^{(1)}\right]
$$
is symmetric under the permutation $i \leftrightarrow j$.
Since also $\left[g^2\check S_i^{(1)}, g^2\check S_j^{(1)}\right] = 0$
and $\left[g^2\check S_i^{(1)}, \check \Pi_j^{(0)}\right] = 0$,
we finally obtain:
$$
\left[\check S_i^{(0)} + g^2\check S_i^{(1)},
\check S_j^{(0)}+ g^2\check S_j^{(1)}\right] = 0,
$$ $$ $$ $$
\left[\check\Pi_i^{(0)}, \check\Pi_j^{(0)}\right] = \left[W(\alpha_i),
W(\alpha_j)\right] = 0,
$$
\begin{eqnarray}
\left[\check S_i^{(0)}  + g^2\check S_i^{(1)}, \check\Pi_j^{(0)}\right] =
\left[\frac{g^2}{\M_i}\check R_i, W(\alpha_j)\right] =
\frac{g^2}{W''(\alpha_i)}
\frac{W'(\alpha_i)-W'(\alpha_j)}{\alpha_{ij}} = g^2\delta_{ij}
\end{eqnarray}
Note that in the last formula there is no need to vary the argument
$\alpha_j$, because this variation gets multiplied by $W'(\alpha_j) = 0$.
When $j = i$ one needs to apply l'Hopital rule to resolve the
$0/0$ ambiguity, and it provides the non-vanishing answer.
This ends the proof of (\ref{quaco}). It is an open question,
if there are higher order corrections $\sim O(g^4)$ to the r.h.s.
of that formula.

Given (\ref{quaco}), of special interest is this kind of quantization of
the WDVV equations.
At least in principle, such approach opens a possibility
to unify WDVV equations for different genera,
the spherical equations from \cite{WDVV,WDVVMMM} with higher genus
equations from \cite{Getz,LoSh}, and thus provide a new deep
connection with geometry of
the moduli space of punctured Riemann surfaces.

\section{Neoclassical period: back to concrete results
and back to physics?}

The story about the challenges of matrix models is nearly infinite
and can be continued further and further. We, however, draw a line
here. The only brief comment that deserves being made at the end
of these notes, is that the {\it transcendental period}, when most
attention is concentrated on non-obvious structures underlying the
subject and thus on high abstractions, is never the final stage of
developement of a theory. Sooner or later, deep insights
from transcendental studies open a
possibility to solve practical problems and, in our case, derive
concrete formulas with the left hand side and the right hand side.
Remarkably, some of such results are already starting to emerge.
Despite they are rather week, it is principally important that they
appear, it means that the theory is on the right
track and new knowledge continues to have many non-trivial
intersections with the old -- what is a necessary feature of a
healthy research project.

We mention here just three results of this kind, all concern the
theory of CIV-DV prepotentials and all establish non-trivial
relations with free fields on Riemann surfaces: the previous-level
theory, from which one starts from when departing into the world of
integrability and matrix models. The first result \cite{Chekhov}
identifies 1-loop correction to the CIV-DV prepotential with
certain combination of determinants in some -- yet
unidentified -- conformal theory. The second result \cite{Kostov2}
provides additional evidence in a form of a CFT representation
(not fully polished yet) of the all-genus matrix-model partition
function $Z_W$.
As expected, this representation actually constructs the
partition function by a Givental-style procedure
from $2n$ ($n = {\rm deg}\ W'(z)$)
Kontsevich $tau$-functions $\tau_K$, obtained by the application of
star operators \cite{Moore} at $2n$ points of the bare spectral
curve (Riemann sphere),  which become ramification points of the
hyperelliptic spectral curve, associated with $Z_W$.
The third result \cite{Eynard} formulates a kind of a perturbation
theory for correlation functions of $2n$ operators inserted at these
points and represents the partition function as a combinatorial
"quantum field theory", essentially of Chern-Simons or, better,
Batalin-Vilkovisky type. We consider this small set of preliminary
results as a clear manifestation that all the ideas, discussed
during the transcendental period can indeed be brought together
and the head can catch the tail: the seemingly transcendental ideas
will finally form a dense network of knowledge, not a shaky road
leading into nowhere...

\section{Acknowledgements}

I am indebted for the invitation, hospitality and support
to L.Baulieu, B.Pioline, E.Rabinovici
and other organizers of Cargese School.
I thank my numerous co-authors and friends and especially
A.Alexandrov, A.Gerasimov, H.Itoyama, S.Kharchev,
A.Marshakov, A.Mironov and A.Zabrodin
for valuable discussions of matrix models.
This work is partly supported by the Federal Program
of the Russian Ministry of
Industry, Science and Technology No 40.052.1.1.1112 and
by the RFBR grant 04-02-16880.

\end{document}